%% file: cartfact.tex
\documentclass[submission,copyright,fleqn]{eptcs}  
\providecommand{\event}{FICS 2015}

\input{packs}
\input{preamble}

\begin{document}

\title{Self-Correlation and Maximum Independence \\ in Finite Relations}

\author{
Dilian Gurov
\institute{KTH Royal Institute of Technology, Stockholm, Sweden}
\email{dilian@csc.kth.se}
\and 
Minko Markov
\institute{ ``St. Kliment Ohridski'' University of Sofia, Sofia, Bulgaria}
\email{minkom@fmi.uni-sofia.bg}}

\def\titlerunning{Self-Correlation and Maximum Independence}
\def\authorrunning{D. Gurov, M. Markov}

\maketitle

\input{abstract}

\input{introduction}

\input{background}

\input{correlation}

\input{independence}

\input{main}

\input{relatedwork}

\input{conclusions}

\input{thebibliography}

\end{document}

%% file: packs.tex
\usepackage{amssymb}
\usepackage{amsmath}

\usepackage{epsfig}
\usepackage{setspace}

\usepackage{xspace}
\usepackage{varioref}
\usepackage{enumerate}
\usepackage{url}
\usepackage{stmaryrd}

\usepackage{relsize}

\usepackage{enumitem}

\usepackage[safe]{tipa}



%% file: preamble.tex
\newtheorem{theorem}{Theorem}
\newtheorem{lemma}{Lemma}
\newtheorem{corollary}{Corollary}
\newtheorem{definition}{Definition}

\newtheorem{proposition}{Proposition}

\newcommand{\BOX}{\ensuremath{\hfill \Box}}

\newcommand{\LA}{\{}
\newcommand{\RA}{\}}
\newcommand{\LAO}{\{}
\newcommand{\RAO}{\}}

\newcommand{\origsqcup}{\sqcup}
\newcommand{\origsqcap}{\sqcap}
\newcommand{\origcup}{\cup}

\newcommand{\origcap}{\cap}

\newcommand{\LDOTS}{\ensuremath{, \allowbreak \ldots, \allowbreak}}
\newcommand{\AB}{\allowbreak}

\newcommand{\newCORR}[2]{\ensuremath{\text{corr}_{{#2}}({#1})}}
\newcommand{\newCOR}[1]{\ensuremath{\text{corr}({#1})}}

\makeatletter

\newcommand{\Rmnum}[1]{\expandafter\@slowromancap\romannumeral #1@}
\makeatother

\newcommand{\restr}[2]{\ensuremath{
  \left.\kern-\nulldelimiterspace 
  #1 
  \vphantom{\big|} 
  \right|_{#2} 
  }}

\makeatletter
\def\@fnsymbol#1{\ensuremath{\ifcase#1\or \dagger\or \ddagger\or
   \mathsection\or \mathparagraph\or \|\or **\or \dagger\dagger
   \or \ddagger\ddagger \else\@ctrerr\fi}}
\makeatother

\newcommand{\RSTR}{\restriction}
\newcommand{\PROJ}[2]{\ensuremath{{#1}\RSTR{#2}}}
\newcommand{\REFINES}{\ensuremath{\sqsubseteq}}
\newcommand{\REFINESproperly}{\ensuremath{\sqsubset}}

\newcommand{\PREC}{\ensuremath{\preccurlyeq}}

\newcommand{\SUBPART}{\ensuremath{\Subset}}

\newcommand{\JOIN}{{\,\mathlarger{\Join}\,}}

\newcommand{\POWERSET}[1]{\ensuremath{\textnormal{\sf \scalebox{.7}[1.0]{POW}}({#1})}}
\newcommand{\POWERSETplus}[1]{\ensuremath{\textnormal{\sf \scalebox{.7}[1.0]{P}}^{+}\!({#1})}}

\newcommand{\ELEMENTS}[1]{\ensuremath{\origcup {#1}}}

\newcommand{\IDEAL}[1]{\mbox{\ensuremath{\downarrow}\ensuremath{#1}}}
\newcommand{\FILTER}[1]{\mbox{\ensuremath{\uparrow}\ensuremath{#1}}}

\newcommand{\PART}[1]{\ensuremath{{\Pi}({#1})}}

\newcommand{\COVER}[1]{\ensuremath{{K}({#1})}}
\newcommand{\INDPART}[2]{\ensuremath{\mathrm{I\Pi}_{#2}({#1})}}
\newcommand{\INDPAR}[1]{\ensuremath{\mathrm{I\Pi}({#1})}}
\newcommand{\MF}[1]{\ensuremath{\textnormal{\sf \scalebox{.7}[1.0]{MF}}\!\left({#1}\right)}}

\newcommand{\MC}[1]{\ensuremath{\textnormal{\sf \scalebox{.7}[1.0]{mincors}}\!\left({#1}\right)}}

\newcommand{\SNGL}[1]{\ensuremath{\textnormal{\sf \scalebox{.7}[1.0]{singletons}}\!\left({#1}\right)}}

\newcommand{\CC}[1]{\ensuremath{\textnormal{\sf \scalebox{.55}[1.0]{CC}}({#1})}}
\newcommand{\CCMF}[1]{\ensuremath%
{\textnormal{\sf \scalebox{.55}[1.0]{CC}}\!\left(\textnormal{\sf \scalebox{.7}[1.0]{MF}}\!\left({#1}\right)\right)}}

\newcommand{\FOC}[1]{\ensuremath{\textnormal{\sf \scalebox{.7}[1.0]{foc}}\!\left({#1}\right)}}

\newcommand{\QUOT}[2]{\ensuremath{%
{\!\left.\raisebox{.08em}{\ensuremath{#1}}/\raisebox{-.12em}{\ensuremath{#2}}\right.}\! }}

\newcommand{\ZERO}{\ensuremath{\bot}\xspace}
\newcommand{\ONE}{\ensuremath{\top}\xspace}

\newcommand{\BOT}{\ensuremath{\bot}\xspace}
\newcommand{\TOP}{\ensuremath{\top}\xspace}

\newcommand{\GLB}{\ensuremath{\origsqcap}\xspace}
\newcommand{\LUB}{\ensuremath{\origsqcup}\xspace}

\newcommand{\mathcall}{\ensuremath{\mathrm}}

%% file: abstract.tex
\begin{abstract}

  We consider relations with no order on their attributes as in
  Database Theory.  An independent partition of the set of
  attributes~$\mathcall{S}$ of a finite relation~$R$ is any partition
  $\mathfrak{X}$ of~$\mathcall{S}$ such that the join of the
  projections of $R$ over the elements of~$\mathfrak{X}$ yields~$R$.
  Identifying independent partitions has many applications and
  corresponds conceptually to revealing orthogonality between sets of
  dimensions in multidimensional point
  spaces. 
  A subset of $\mathcall{S}$ is termed self-correlated if there is a value of
  each of its attributes such that no tuple of $R$ contains all those
  values.  This paper uncovers a connection between independence and
  self-correlation, showing that the maximum independent partition is
  the least fixed point of a certain inflationary transformer $\alpha$
  that operates on the finite lattice of partitions of $\mathcall{S}$.
  $\alpha$ is defined via the minimal self-correlated subsets of
  $\mathcall{S}$.
  We use some additional properties of $\alpha$ to show the said fixed
  point is still the limit of the standard approximation sequence,
  just as in Kleene's well-known fixed point theorem for continuous
  functions.  

\end{abstract}

%% file: introduction.tex
\section{Introduction} 

The problem of discovering independence between sets of points in a
multidimensional space is a fundamental problem in science. It arises
naturally in many areas of Computer Science.  
For instance, with respect to relational data,
discovering such independence allows exponential gains in storage
space and processing of information \cite{OZ}, \cite{AKO09}, and can
facilitate the problem of machine learning \cite{Rendle}.  With
respect to problem clusterisation of multidimensional relational data,
finding independence helps finding the desired clusters
\cite{Gunnemann}, \cite{Lin}.  Decomposing data into smaller units
that are independent except at their interfaces has been known
to be essential for understanding large legacy systems \cite{WHH}.
Independence has also been the subject of recent works in logic, giving 
rise to so-called logics of dependence and 
independence~\cite{Gra-Vaa-13-Stud-Log}. 

The concrete motivation for the present work derives from the area of
\emph{software product line engineering}, a discipline that aims at
planning for and developing a \emph{family} of products through
managed reuse in order to decrease time to market and improve software
quality~\cite{Pohl}.  A software family can be modelled as a relation
whose attributes are the software's functionalities. The various
implementations of each functionality in the form of software
artefacts are the attributes' \emph{values}. The individual products
of a family are thus modelled as the tuples of that relation over the
attributes.  In previous works
\cite{GurOstSchae11ISoLA,SchGurSol11FMCO} we considered a restricted
class of software families called \emph{simple families} (later on we
changed the term ``families'' to the more abstract term
``relations''), where discovery of independence and a compositional
model checking technique are utilised to derive a
\emph{divide-and-conquer verification strategy}.  Simple relations
constitute the least class that contains the single-attribute,
single-value relations and is closed under join of relations with
disjoint attribute sets and unions of relations over the same set of
attribute names but with disjoint value sets.  In the present work we
generalise these previous results to discovering independence in
arbitrary relations.  We investigate decompositions of a relation~$R$
with disjoint attributes such that $R$ equals the join of the
component relations.  Every decomposition is represented by a
partition of the set of attributes of~$R$.  Such partitions are termed
\emph{independent partitions}.

The problem of computing a maximum decomposition of this kind has
previously been studied in~\cite{AKO}, where it is referred to as
\emph{prime factorisation}, and an efficient algorithmic solution is
proposed.  In this paper we investigate an alternative approach that
works purely on the level of the attributes of~$R$ and is based on the
concept of \emph{correlation} between attributes.  We have discovered
a nontrivial connection between independence and correlation and the
major goal of this paper is to demonstrate that connection.

A first observation is that the decomposition problem cannot be solved
purely based on analysis of pairs of attributes.  In the
aforementioned work~\cite{GurOstSchae11ISoLA} we compute dependence
(or independence) in simple relations by computing correlation between
pairs of attributes.  That approach does not generalise for arbitrary
relations as we show in this paper.  Our solution is to introduce
\emph{self-correlation} of sets (of arbitrary cardinality) of
attributes.  In other words, the current notion of correlation is a
hypergraph whose hyperedges are the self-correlated sets, rather than
an ordinary graph as were the case with the simple relations.  Since
self-correlated sets are upward closed under set inclusion
(Proposition~\ref{obs1}), the minimal self-correlated sets, or the
\emph{mincors} (Definition~\ref{def4}), are the foundation of our
analysis.  A second observation is that mincors do not cross
independent partitions (Lemma~\ref{lemma2}), hence one can safely
merge overlapping mincors to compute the maximum independent
partition. In the case of simple relations that merger indeed yields
the maximum independent partition \cite{GurOstSchae11ISoLA} but in
arbitrary relations merging the mincors \emph{does not} necessarily
output an independent partition, as the example \vpageref{example1}
shows.  We overcome this hindrance with the help of a final important
insight.  Let $\mathfrak{X}$ be the partition of the set of attributes
that results from merging overlapping mincors.  The relation can be
factored on $\mathfrak{X}$, producing a \emph{quotient relation}.  In
other words, the elements of $\mathfrak{X}$ are considered atomic now;
the subsets of $\mathfrak{X}$ may or may not be self-correlated in
their turn, and the said quotient relation is defined via those new
mincors.
We show that the procedure of identifying mincors and
merging overlapping ones can be repeated on this quotient relation and
this can be iterated until stabilisation, yielding the desired maximum
independent partition.

The above insights suggest that relational decomposition can be
presented in terms of a transformer over the finite lattice of
quotient relations, or conceptually even simpler, over \emph{the
  lattice of the partitions} ordered by refinement, inducing the
former lattice.
The transformer~$\alpha$ on partitions introduced here essentially 
corresponds to identifying the mincors of the quotient relation 
induced by a partition, merging the overlapping ones, and extracting
from the result the corresponding partition (Definition~\ref{def:alpha}).
We prove that the independent partitions correspond exactly to the fixed
points of~$\alpha$ (Theorem~\ref{theorem:alphaindependent}).

If~$\alpha$ is monotone, one can utilise two well-known fixed point
theorems on complete lattices (having in mind that monotone functions
over finite lattices are continuous).
First, by Tarski's fixed point theorem for complete
lattices~\cite{Tarski}, the set of fixed points forms a lattice itself
with respect to the same ordering, hence there is a unique \emph{least
  fixed point} (LFP), which in our case would be precisely the maximum
independent partitioning that we are after.
And second, one can utilise Kleene's fixed point theorem~\cite{LNS}, 
to the effect that the LFP can be computed
\emph{iteratively}, starting from the bottom of the lattice,
\emph{i.e.} the partition into singletons, and applying~$\alpha$ until
stabilisation, \emph{i.e.}, until the fixed point is reached.
It turns out, however, that~$\alpha$  in general is \emph{not monotone}
as demonstrated by the example \vpageref{alpha-not-monotone}  
and therefore the above reasoning is not applicable. 

On the other hand, we show that~$\alpha$ is \emph{inflationary}
(Proposition~\ref{prop3}).  The existence of a LFP is established by
showing that there exists a fixed point and the set of all fixed
points is closed under intersection
(Lemma~\ref{lemma:meetindependent}). Furthermore, the downward closure
of LFP, \emph{i.e.}, the set of all partitions refining it, is closed
under~$\alpha$ (Lemma~\ref{lemma:idealclosed}).  Since the lattice is
finite, these results give rise to a modified version of Kleene's
fixed point theorem---formulated in terms of inflationary transformers
rather than monotone ones (Theorem~\ref{theorem:main})---justifying
the same iterative fixed point computation procedure
(Corollary~\ref{cor:fixed-point-comp}).  The proposed characterisation
reduces relational decomposition to the problem of identifying the
mincors of a relation.

\paragraph{Organisation}
The paper is organised as follows. %
Section~\ref{sec:background} recalls some known notions and results
about sets and families, partitions, lattices, fixed points,
relations, attributes, and relation schemes, quotient relations, and
defines independent partitions of the attributes set. %
Section~\ref{sec:correlation} develops the theory of self-correlated
sets in quotient relations and how they relate w.r.t. partition
abstraction.  %
Section~\ref{sec:independence} presents many useful lemmas that
concern independence. %
Section~\ref{sec:main} defines the transformer $\alpha$ and contains
our main result, Theorem~\ref{theorem:main}. %
Section~\ref{sec:relatedwork} discusses what we currently know about
the area of decomposition of relations, also called factorisation of
relations, and compares the approach and the results of this paper
with similar works. %
The final Section~\ref{sec:conclusion} draws some conclusions and
outlines directions for future work.

%% file: background.tex
\section{Background}
\label{sec:background}

In this section we recall some standard set-theoretical notions and 
notation needed for our theoretical developments.

\subsection{Sets, covers, and partitions}

In this work we consider only finite sets.  The powerset of a set $A$
is denoted by \POWERSET{A} and \POWERSETplus{A} denotes $\POWERSET{A}
\setminus \{ \emptyset \}$.  \emph{Ground sets} are nonempty sets over
which we construct the families that are our subject of research.

Let $A$ be a ground set.  \emph{A family over $A$} is any nonempty
subset of \POWERSETplus{A}.  
A family $\mathcall{F}$ is \emph{Sperner family} if $\forall X, Y \in
\mathcall{F}: X \not\subseteq Y$.  $\mathcall{F}$ is \emph{connected} if
$\forall X, Z \in \mathcall{F}$: $X \cap Z \not= \emptyset$ or
$\mathcall{F}$ has elements $Y_1$, $Y_2$, \ldots, $Y_k$ for some $k
\geq 1$, such that $X \cap Y_1 \not= \emptyset$, $Y_i \cap Y_{i+1}
\not= \emptyset$ for $1 \leq i \leq k-1$, and $Y_k \cap Z \not=
\emptyset$.  \emph{A connected component of a family} is any maximal
connected subfamily in it.  
We use \CC{\mathcall{F}} to denote the family
$\{\ELEMENTS{\mathcall{B}} \, | \, \mathcall{B} \text{ is a connected
  component of } \mathcall{F} \}$.  \emph{A superfamily over $A$} is
any nonempty subset of \POWERSETplus{\POWERSETplus{A}}.

Suppose $A$ is a set.  \emph{A cover of $A$} is any family
$\mathcall{F}$ over $A$ such that $\origcup \mathcall{F} = A$.  The set
of all covers of~$A$ is denoted by~\COVER{A}.  If $\mathfrak{X} \in
\COVER{A}$ and $Y \cap Z = \emptyset$ for all distinct $Y, Z \in
\mathfrak{X}$, we say $\mathfrak{X}$ is \emph{a partition of}~$A$.  If
$|\mathfrak{X}| = 1$ the partition is \emph{trivial} and if
$|\mathfrak{X}| = |A|$ the partition is \emph{partition into
  singletons}.  Note that \CC{\mathcall{F}} defined above is a
partition of the ground set.  We denote by $\mathfrak{Y} \SUBPART
\mathfrak{X}$ the fact that for some $B \subseteq A$, $\mathfrak{Y}$ is
a family over $B$ such that every element of $\mathfrak{Y}$ is a
subset of precisely one element of $\mathfrak{X}$ and every element of
$\mathfrak{X}$ is a superset of at most one element of $\mathfrak{Y}$.
For example, if $A = \{a,b,c,d,e,f, g, h, k\}$ then $\{ \{b\}, \{c\},
\{d,g\} \} \SUBPART \{ \{a,b\}, \{c\}, \{d,e,f,g\}, \{h,k\}\}$.

The set of all partitions of $A$ is denoted by \PART{A}.  For any
$P_1, P_2 \in \PART{A}$, \emph{$P_1$ refines $P_2$}, which we denote
by $P_1 \REFINES P_2$, if
\[
\forall X \in P_1 \;\; \exists Y \in P_2 : X \subseteq Y
\]
Conversely, we say that \emph{$P_2$ abstracts $P_1$}.  If $P_1
\REFINES P_2$ and $P_1 \not= P_2$ we write $P_1 \REFINESproperly P_2$.

\subsection{Partial orders, lattices, and chains}

We denote generic partial orders by ``\PREC''.  If $(A, \PREC)$ is a
poset, \emph{a least element} of~$A$ is any $x \in A$ such that
$\forall y \in A: x \PREC y$ and \emph{a greatest element} of~$A$ is
any $x \in A$ such that $\forall y \in A: y \PREC x$.  A least element
may not exist but if it exists it is unique; the same holds for a
greatest element.  The least element is called \emph{bottom} and is
denoted by \BOT.  The greatest element is called \emph{top} and is
denoted by \TOP.  \emph{A chain} in a poset $(A, \PREC)$ is any $B
\subseteq A$ such that $\forall x,y \in B: x \PREC y \vee y \PREC x$.

\emph{A lattice} is a poset $(A, \PREC)$, shortly $A$ when $\PREC$ is
understood, such that for any $x, y \in A$ there exists a (unique)
greatest lower bound in $A$ called \emph{meet} and denoted by $x \GLB
y$ and a (unique) least upper bound in $A$ called \emph{join} and
denoted by $x \LUB y$.  Collectively, $\GLB$ and $\LUB$ are \emph{the
  lattice operations of $A$}.  They are commutative and associative
\cite[pp.\ 8]{Birkhoff}.  We generalise the lattice operations on
subsets of $A$ in the obvious way.  \emph{A complete lattice} is a
lattice such that every $B \subseteq A$ has a meet $\GLB B$ and a join
$\LUB B$.  In particular, $A$~has a meet $\GLB A = \BOT$ and a join
$\LUB A = \TOP$.  Every finite lattice is
complete 
\cite[pp.\ 46]{DaveyPriestley}, therefore from now on by lattice we
mean complete lattice.
For any $x \in A$, the sets $\{y \in A \, | \, y \PREC x \}$ and $\{y
\in A \, | \, x \PREC y \}$ are called \emph{down-$x$} and
\emph{up-$x$} and are denoted by \FILTER{x} and \IDEAL{x},
respectively \cite[pp.\ 20]{DaveyPriestley}.

It is well-known that $(\PART{A}, \REFINES)$ is a lattice.
Furthermore, \ZERO is the partition into singletons, \ONE is the
trivial partition, and for any $P_1, P_2 \in \PART{A}$, $P_1 \GLB P_2
= \{X \cap Y \, | \, X \in P_1, Y \in P_2 \} \setminus \{ \emptyset
\}$ and $P_1 \LUB P_2 = \CC{P_1 \cup P_2}$ (see
\cite[pp. 15]{Birkhoff}).  We extend the ``$\GLB$'' notation to
subsets of partitions: for any $\mathfrak{X}, \mathfrak{Y}
\in \PART{A}$, for any nonempty $\mathfrak{X}' \subseteq \mathfrak{X}$
and any nonempty $\mathfrak{Y}' \subseteq \mathfrak{Y}$ such that
$\mathfrak{X}' \cap \mathfrak{Y}' \not= \emptyset$, $\mathfrak{X}' \GLB
\mathfrak{Y}'$ denotes the set $\{B \cap C \, | \, B \in \mathfrak{X}',
C \in \mathfrak{Y}'\} \setminus \{\emptyset \}$.



\subsection{Functions and fixed points}

Suppose $A$ is a set and $f: A \rightarrow A$ is a function.  For
every $x \in A$: $f^0(x) \stackrel{\text{def}}{=} x$ and for every $n
\in \mathbb{N}^+$, $f^n(x) \stackrel{\text{def}}{=} f \circ
f^{n-1}(x)$.  For every $n \in \mathbb{N}$, $f^n(x)$ is \emph{the
  $n$-th iterate of $f$}.  \emph{A fixed point} of $f$ is every $x \in
A$ such that $f(x) = x$.  Let $(A, \PREC)$ be a poset.  A function $f:
A \rightarrow A$ is \emph{monotone} if $\forall x, y \in A: x \PREC y
\rightarrow f(x) \PREC f(y)$ and
$f$~is \emph{inflationary} if $\forall x \in A: x \PREC f(x)$
\cite[pp. 263]{Roman}.

A well-known fixed point theorem is Tarski's fixed point theorem for 
continuous functions over complete lattices~\cite{Tarski}, stating 
that the set of fixed points is non-empty and forms a lattice itself
with respect to the same ordering, and hence the function has a unique 
\emph{least fixed point} (LFP). 
Another well-known theorem due to Kleene states the existence of an 
LFP for continuous functions on chain-complete partial orders~\cite{LNS},
and that the LFP can be computed \emph{iteratively}, starting from the 
bottom of the lattice and applying the function until stabilisation.


\subsection{Schemes, relations, and quotient relations}

The following definitions are close to the ones in \cite{Maier}.
\emph{A scheme} is a nonempty set $\mathcall{S} = \{A_1, \ldots, A_n
\}$ whose elements, called \emph{the attributes}, are nonempty sets.
For every attribute, its elements are said to be its \emph{ values}.
\emph{A relation over $S$} is a nonempty set of total functions
$\{t_1, t_2 \LDOTS t_p\}$, which we call \emph{the tuples}, such that
for $1 \leq j \leq p$, $t_j: \mathcall{S} \rightarrow
\ELEMENTS{\mathcall{S}}$, with the restriction that $t_j(A_i) \in
A_i$, for $1 \leq i \leq n$.  We assume that every value of every
attribute occurs in at least one tuple.

The relations we have in mind are as in Relational Database Theory,
\emph{i.e.} with unordered tuples, rather than as in Set Theory,
\emph{i.e.} with ordered tuples.

We further postulate that the said attributes are mutually disjoint
sets.  That allows a simplification of the definition of relation: a
relation over $\mathcall{S}$ is nonempty set of tuples, each tuple
being an $n$-element set with precisely one element from every
attribute.  To save space, we often write the tuples without commas
between their elements.  For example, let $n= 3$, $A_1 = \{a_1,
a_2\}$, $A_2 = \{b_1, b_2\}$, and $A_3 = \{c_1, c_2, c_3\}$.  One of
the relations over the scheme $\{A_1, A_2, A_3\}$ is written as $\{\LA
a_1 b_1 c_1\RA, \LA a_1 b_2 c_2\RA, \LA a_2 b_2 c_3\RA \}$.

Let $\mathcall{S}_1, \mathcall{S}_2, \ldots, \mathcall{S}_k$ be
schemes such that for $1 \leq i < j \leq k$, $\forall A \in
\mathcall{S}_i \; \forall B \in \mathcall{S}_j: A \cap B = \emptyset$.
Let $R_i$ be a relation over $\mathcall{S}_i$, for $1 \leq i \leq k$.
\emph{The join of $R_1$, \ldots, $R_k$} \label{bowtie} is the relation
\[
R_1 \Join R_2 \Join \cdots \Join R_k = \{ \ELEMENTS{\LA x_1, x_2, \ldots, x_k \RA} \, | \, x_1 \in R_1, x_2 \in R_2, \ldots, x_k \in R_k \}
\]

\emph{The complete relation} \label{complete-relation} over
$\mathcall{S} = \{A_1, \ldots, A_n \}$ is $\Join_{i=1}^n \{ \{x\} \, | \, x \in A_i  \}$.
Clearly, its cardinality is $\prod_{i=1}^n |A_i|$.

Let $\mathcall{S} = \{A_1, \ldots, A_n \}$ be a scheme.  \emph{A
  subscheme of $\mathcall{S}$} is any nonempty subset of
$\mathcall{S}$.  The notation \restr{f}{Z} stands for the restriction
of $f$ to $Z$, for any function $f: X \rightarrow Y$ and any $Z
\subseteq X$.  Let $R = \{t_1, t_2, \ldots, t_p\}$ be a relation over
$\mathcall{S}$ and let $\mathcall{T}$ be a subscheme of
$\mathcall{S}$.  \emph{The projection of $R$ on $\mathcall{T}$} is
$\PROJ{R}{\mathcall{T}} = \{\restr{t_j}{\mathcall{T}} : 1 \leq j \leq
p\}$. 


\begin{definition}[quotient relation]
  \label{label:quotrel}
  Let $R$ be a relation over some scheme $\mathcall{S}$.  For any
  $\mathfrak{X} = \{\mathcall{X}_1, \AB \mathcall{X}_2 \LDOTS
  \mathcall{X}_n\} \in \PART{\mathcall{S}}$, $\QUOT{R}{\mathfrak{X}}
  \subseteq \Join_{i=1}^n (\PROJ{R}{\mathcall{X}_i})$ is the
  following relation:
  \begin{alignat*}{1} 
      & \forall \LA y_1 y_2 \ldots y_n \RA \in  {\Join_{i=1}^n}
      (\PROJ{R}{\mathcall{X}_i}): \\
      & \qquad \qquad \qquad  \LA y_1 y_2 \ldots y_n \RA \in \QUOT{R}{\mathfrak{X}} \; \;
      \mathrm{ iff } \; \; \exists t \in R \; \forall i_{\, 1 \leq i
        \leq n \,} ( \PROJ{t}{\mathcall{X}_i} = y_i )
  \end{alignat*}
\noindent
We term \QUOT{R}{\mathfrak{X}} the \emph{quotient relation of~$R$
relative to $\mathfrak{X}$}.  When $\mathfrak{X}$ is understood we
say simply \emph{the quotient relation of~$R$}.
\end{definition}
\noindent We emphasise the quotient relation is not over $\mathcall{S}$
but over a partition of $\mathcall{S}$.
 
\medskip
\noindent
Here is an example of a quotient relation.  Let $\mathcall{S} = \{A,
\AB B, \AB C, \AB D\}$, let each attribute have precisely two values,
say $A = \{a_1, a_2\}$ and so on, let $\mathfrak{X}_1 = \{ \{A, B\},
\{C , D\} \}$, let $\mathfrak{X}_2 = \{ \{A\}, \{B\}, \{C\}, \{D\} \}$,
and let
\begin{equation}
  R' = 
  \{  \LA a_1 b_1  c_1 d_1 \RA, \LA a_1 b_1 c_2 d_2 \RA,
  \LA a_1 b_2 c_1 d_2 \RA,
  \LA a_2 b_2 c_1 d_1 \RA,  \LA a_2 b_2 c_2 d_2\RA  \} 
  \label{eq3} 
\end{equation}
be a relation over $\mathcall{S}$.  Then
\begin{alignat}{2}
 \QUOT{R'}{\mathfrak{X}_1} & = \{ & & \LA \LAO a_1, b_1 \RAO  \LAO c_1, d_1 \RAO  \RA, 
                                                      \LA \LAO a_1, b_1 \RAO  \LAO c_2, d_2 \RAO \RA, 
                                                      \LA \LAO a_1, b_2 \RAO  \LAO c_1, d_2 \RAO \RA, \notag \\ 
                                                     &&& \LA \LAO a_2, b_2 \RAO  \LAO c_1, d_1 \RAO \RA, 
                                                      \LA \LAO a_2, b_2 \RAO  \LAO c_2, d_2 \RAO \RA  \} \label{eq6}
                                                      \\
 \QUOT{R'}{\mathfrak{X}_2} & = \{ && 
                                                                   \LA 
                                                                       \LAO a_1 \RAO  
                                                                       \LAO  b_1 \RAO 
                                                                       \LAO c_1 \RAO 
                                                                       \LAO d_1 \RAO  
                                                                   \RA, 
                                                                   \LA 
                                                                       \LAO a_1 \RAO  
                                                                       \LAO  b_1 \RAO 
                                                                       \LAO c_2 \RAO 
                                                                       \LAO d_2 \RAO  
                                                                   \RA, 
                                                                   \LA 
                                                                       \LAO a_1 \RAO  
                                                                       \LAO  b_2 \RAO 
                                                                       \LAO c_1 \RAO 
                                                                       \LAO d_2 \RAO  
                                                                   \RA, \notag \\ 
                                                      &&& 
                                                                   \LA 
                                                                       \LAO a_2 \RAO  
                                                                       \LAO  b_2 \RAO 
                                                                       \LAO c_1 \RAO 
                                                                       \LAO d_1 \RAO  
                                                                   \RA, 
                                                                   \LA 
                                                                       \LAO a_2 \RAO  
                                                                       \LAO  b_2 \RAO 
                                                                       \LAO c_2 \RAO 
                                                                       \LAO d_2 \RAO  
                                                                   \RA \label{eq7} 
  \}
\end{alignat}
A quotient relation is but a grouping together of the tuples of the
original relation into subtuples according to the partition.  It
trivially follows that $|\!\QUOT{R}{\mathfrak{X}}\!| = |R|$ for any
relation $R$ over any attribute set $\mathcall{S}$ and any
$\mathfrak{X} \in
\PART{\mathcall{S}}$.

\subsection{Independent partitions}


For a given relation $R$ over some scheme $S$, we are after decompositions of
$R$ such that $R$ equals the join of the obtained components.  Each
decomposition of this kind corresponds to a certain partition of $S$.
\begin{definition}[independent partition]
  \label{def:IP}
  Let $R$ be a relation over some scheme $S$.  For any $\mathfrak{X}
  \in \PART{\mathcall{S}}$, $\mathfrak{X}$~is \emph{an independent
    partition of $\mathcall{S}$ with respect to $R$} if $\displaystyle
  R = \underset{{Y} \in \mathfrak{X}}{\JOIN} \, \PROJ{R}{{Y}}$.  The
  set of all independent partitions of $\mathcall{S}$ with respect to
  $R$ is denoted by \INDPART{\mathcall{S}}{R}, or shortly
  \INDPAR{\mathcall{S}} if $R$ is understood.  If a partition is not
  independent, it is \emph{dependent}.
\end{definition}
Note that \INDPAR{\mathcall{S}} is nonempty since it necessarily
contains the trivial partition.

\begin{proposition} 
  \label{prop2}
  For every independent partition $\mathfrak{X}$,
  $\QUOT{R}{\mathfrak{X}}$ is the complete relation over
  $\mathfrak{X}$.
\end{proposition}

Informally speaking, the object of the present study is the
independent partition with the maximum number of equivalence classes,
provided it is unique.

%% file: correlation.tex
\section{Correlation in Relations}
\label{sec:correlation}

In this section we define correlation in relations and quotient relations.
From now on assume an arbitrary but fixed scheme~$\mathcall{S}$ and 
relation~$R$ over it.

\subsection{Correlated subsets of ground sets}
\label{subsection:corr}

In this subsection, the ground sets are schemes.
\begin{definition}[correlated subsets of schemes]
  \label{def3} 
  Let $\mathcall{S} = \{A_1, A_2 \LDOTS A_n\}$ and let $\mathcall{T}$
  be some nonempty subscheme $\{A_{i_1}, A_{i_2} \LDOTS A_{i_m}\}$
  where $1 \leq i_1 < i_2 < \cdots < i_m \leq n$.  $\mathcall{T}$~is
  \emph{self-correlated with respect to $R$}, or shortly
  \emph{correlated with respect to $R$}, iff
  \begin{equation}
  \label{eq1}
    \exists x_1 \in A_{i_1} \;
    \exists x_2 \in A_{i_2} \; 
    \cdots \;  
    \exists x_m \in A_{i_m} :
    \LA x_1 x_2 \cdots x_m \RA \not\in \PROJ{R}{\mathcall{T}}
  \end{equation}
  We denote that fact by $\newCORR{\mathcall{T}}{R}$ or
  $\newCOR{\mathcall{T}}$ if $R$ is understood.  The opposite concept
  is \emph{uncorrelated}.  The family $\{\mathcall{T} \subseteq
  \mathcall{A} \, | \, \newCORR{\mathcall{T}}{R} \}$, in case it is nonempty, is
  called \emph{the correlation family of $R$}.
\end{definition}
\noindent
Note that no minimal correlated subset is a singleton.
%
The following result re-states correlation of a subscheme in terms of the
projection of the relation on it.
\begin{lemma}
  \label{lemma1}
  Let $\mathcall{T} \subseteq \mathcall{S}$.  Then
  $\newCOR{\mathcall{T}}$ iff $\PROJ{R}{\mathcall{T}} \subsetneq
  {\JOIN}_{{X \in \mathcall{T}}} \, \PROJ{R}{\{X\}}$.
\end{lemma}

\noindent
\emph{Proof:} \,
  First assume $\newCOR{\mathcall{T}}$.  By
  Definition~\ref{def3}, there is an element in every attribute from
  $\mathcall{T}$ such that the tuple of those elements does not occur
  in \PROJ{R}{\mathcall{T}}.  On the other hand, the tuples of
  ${\JOIN}_{{X \in \mathcall{T}}} \,
  \PROJ{R}{\{X\}}$ are all possible combinations of the elements of
  the attributes in $\mathcall{T}$.  Therefore, $\PROJ{R}{\mathcall{T}}
  \subsetneq \, {\JOIN}_{{X \in \mathcall{T}}} \, \PROJ{R}{\{X\}}$.

  In the other direction, assume $\neg \newCOR{\mathcall{T}}$.  The
  negation of expression (\ref{eq1}) in Definition~\ref{def3} is but
  another way to write $\PROJ{R}{\mathcall{T}} = \,
  {\JOIN}_{{X \in \mathcall{T}}} \, \PROJ{R}{\{X\}}$.  \BOX

As the next result establishes, with respect to the poset 
$(\mathcall{S}, \subseteq)$, every correlated subset is upward closed,
while every uncorrelated subset is downward closed.
\begin{proposition}
  \label{obs1} If $\newCOR{\mathcall{T}}$ for some $\mathcall{T}
  \subseteq \mathcall{S}$ then $\forall \mathcall{Z}_{\, \mathcall{T}
    \subseteq \mathcall{Z} \subseteq \mathcall{S}}:
  \newCOR{\mathcall{Z}}$.  If $\neg \newCOR{\mathcall{T}}$ for some
  $\mathcall{T} \subseteq \mathcall{S}$ then $\forall \mathcall{Z}_{\,
    \mathcall{Z} \subseteq \mathcall{T}}: \neg \newCOR{\mathcall{Z}}$.
\end{proposition}
It is obvious that the correlation family, if it exists, is a cover of
the scheme.  Furthermore, it does not exist iff the relation is
complete.  The interesting part of a correlation family is the
sub-family comprising the minimal correlated sets.  However, that
sub-family does not necessarily cover the scheme.  We want to define a
family that both covers the scheme---because we are ultimately
interested in a partition of the scheme---and is a Sperner family,
since the implied members of the family are of no interest.
\begin{definition}[mincor family]
  \label{def4}
  \emph{A mincor of $R$} is every minimal, self-correlated with
  respect to $R$, subscheme $\mathcall{T} \subseteq \mathcall{S}$.
  Further, $\MC{R} \stackrel{\text{def}}{=} \{ \mathcall{T} \subseteq
  \mathcall{S} \, | \, \mathcall{T} \text{ is a mincor} \,\}$ and
  $\SNGL{R} \stackrel{\text{def}}{=} \{ \{ A \} \, | \, A \in
  \mathcall{S} \wedge \neg \exists X \in \MC{R} : A \in X \}$.
  \emph{The mincor family} of $R$, denoted by \MF{R}, is $ \MF{R} =
  \MC{R} \cup \SNGL{R}$.
\end{definition}
For example, \label{example2} consider $R'$ defined in (\ref{eq3})
\vpageref{eq3}.  Clearly, $\newCORR{\{A, B\}}{R'}$ and $\newCORR{\{C,
  D\}}{R'}$ because of the lacks of both $a_2$ and $b_1$ in any tuple
and the lack of both $c_2$ and $d_1$ in any tuple, respectively.  The
other four two-element subsets of $\mathcall{S}$ are uncorrelated.
Then $\SNGL{R'} = \emptyset$ and therefore $\MF{R'} = \{ \{A, B\},
\{C, D\} \}$.

\begin{proposition}
  \label{obs2}
  With respect to $\mathcall{S}$ and $R$, \MF{R} exists and is
  unique.  
\end{proposition}
If $R$~is complete then \MF{R}~consists of singletons.  
Clearly, $\MF{R} \in \COVER{S}$, and thus $\CCMF{R} \in \PART{S}$.

\subsection{Correlation in quotient relations}

The following result establishes an important connection between
self-correlation in a partition of the scheme and self-correlation in
the scheme itself.  More specifically, Lemma~\ref{lemma:twoways} is
used to prove Lemma~\ref{lemma:mincors}, and the latter is
used in the proof of Lemma~\ref{lemma4} \vpageref{lemma4}.
\begin{lemma}
  \label{lemma:twoways}
  For any $\mathfrak{X} \in \PART{\mathcall{S}}$ and
  $\mathfrak{X}' \subseteq \mathfrak{X}$:
  \[
  \newCORR{\mathfrak{X}'}{\scriptsize \QUOT{R}{\mathfrak{X}}} \leftrightarrow
  \newCORR{\ELEMENTS{\mathfrak{X}'}}{R}
  \]
\end{lemma}

\noindent
\emph{Proof:} \,
  Assume \newCORR{\mathfrak{X}'}{\scriptsize
    \QUOT{R}{\mathfrak{X}}}.  Let $\mathfrak{X}' = \{Y_1, Y_2 \LDOTS
  Y_m \}$.  So, \PROJ{(\QUOT{R}{\mathfrak{X}})}{\mathfrak{X}'} does
  not contain some $m$-tuple $\LA U_1, U_2 \LDOTS U_m \RA$ such that
  $U_i \in \PROJ{R}{Y_i}$ for $1 \leq i \leq m$.  Then
  \PROJ{R}{\ELEMENTS{\mathfrak{X}'}} does not contain
  \ELEMENTS{\LA U_1, U_2 \LDOTS U_m \RA}.

  In the other direction, assume \newCORR{\ELEMENTS{\mathfrak{X}'}}{R}
  where \ELEMENTS{\mathfrak{X}'} is a subset $\mathcall{S}'$ of
  $\mathcall{S}$.  Let $\mathcall{S}' = \{A_1, A_2 \LDOTS A_n\}$.  That
  is, \PROJ{R}{\mathcall{S}'} does not contain some $n$-tuple $\LA W_1,
  W_2 \LDOTS W_n \RA$ such that $W_i \in A_i$ for $1 \leq i \leq n$.
  Let $\mathfrak{X}' = \{Y_1, Y_2 \LDOTS Y_m \}$.  Then
  \PROJ{(\QUOT{R}{\mathfrak{X}})}{\mathfrak{X}'} does not contain the
  $m$-tuple $\LA U_1, U_2 \LDOTS U_m\RA$ where $U_i \in \PROJ{R}{Y_i}$
  for $1 \leq i \leq m$.  
\BOX

\noindent
As an example that illustrates Lemma~\ref{lemma:twoways}, consider
$R'$ and $\mathfrak{X}_1$ \vpageref{eq3}.  Clearly, $\mathfrak{X}_1 =
\{\{A, B\}, \{C , D\}\}$ is self-correlated with respect to
\QUOT{\tilde{R}}{\mathfrak{X}_1} as \QUOT{\tilde{R}}{\mathfrak{X}_1}
does not contain, among others, the tuple $\LA \LAO a_1, b_1 \RAO \LAO
c_1, d_2 \RAO \RA$.  That implies $\ELEMENTS{\mathfrak{X}_1} =
\{A,B,C,D\}$ is self-correlated with respect to $\tilde{R}$: since
$\LA \LAO a_1, b_1 \RAO \LAO c_1, d_2 \RAO \RA$ is not an element of
\QUOT{\tilde{R}}{\mathfrak{X}_1}, it must be the case that $\LA a_1
b_1 c_1 d_2 \RA$ is not element of $\tilde{R}$ (and indeed it is not).
In the other direction, the fact that $\LA a_1 b_1 c_1 d_2 \RA \not\in
\tilde{R}$ implies $\LA \LAO a_1, b_1 \RAO \LAO c_1, d_2 \RAO \RA
\not\in \QUOT{\tilde{R}}{\mathfrak{X}_1}$.

\medskip
\noindent
The next result establishes that for every mincor~$\mathcall{Y}$ of a 
quotient relation there is a way to pick elements from every element 
of~$\mathcall{Y}$ such that the collection of those elements is a mincor 
of the original relation~$R$.
\begin{lemma}
\label{lemma:mincors}
   $\forall \mathfrak{X} \in \PART{\mathcall{S}} \; \forall \mathfrak{Y}  \in \MC{\QUOT{R}{\mathfrak{X}}} \;
  \exists \mathcall{Z} \SUBPART \mathfrak{Y}: |\mathcall{Z}|
  = |\mathfrak{Y}| \wedge \ELEMENTS{\mathcall{Z}} \in \MC{R}$.
\end{lemma}

\noindent
\emph{Proof:} \,
  Assume $\mathfrak{Y} \in \MC{\QUOT{R}{\mathfrak{X}}}$.  Clearly,
  there is some $\mathcall{Z} \SUBPART \mathfrak{Y}$ such that
  \ELEMENTS{\mathcall{Z}} is correlated with respect to $R$ because
  $\SUBPART$ is reflexive and \ELEMENTS{\mathfrak{Y}} is correlated
  with respect to $R$ by Lemma~\ref{lemma:twoways}.  Now consider any
  $\mathcall{Z}' \SUBPART \mathfrak{Y}$ such that $|\mathcall{Z}'| <
  |\mathfrak{Y}|$.  There exists some $\mathfrak{Y}' \subset
  \mathfrak{Y}$ such that $\mathcall{Z} \SUBPART \mathfrak{Y}'$.  But
  $\mathfrak{Y}'$ is uncorrelated with respect to
  \QUOT{R}{\mathfrak{X}} because $\mathfrak{Y}$ is a mincor of
  \QUOT{R}{\mathfrak{X}} and so every proper subset of $\mathfrak{Y}$
  is uncorrelated with respect to \QUOT{R}{\mathfrak{X}}.  Note that
  $\mathfrak{Y}'$ being uncorrelated with respect to
  \QUOT{R}{\mathfrak{X}} implies \ELEMENTS{\mathcall{Z}'} is
  uncorrelated with respect to $R$ by Lemma~\ref{lemma:twoways}.  It
  follows that for any $\mathcall{Z} \SUBPART \mathfrak{Y}$ such that
  \newCORR{\ELEMENTS{\mathcall{Z}}}{R}---and we established such a
  $\mathcall{Z}$ exists---it is the case that $|\mathcall{Z}| =
  |\mathfrak{Y}|$.

  So, there exists a $\mathcall{Z} \SUBPART \mathfrak{Y}$ such that
  $|\mathcall{Z}| = |\mathfrak{Y}|$ and \ELEMENTS{\mathcall{Z}} is
  correlated with respect to $R$.  Furthermore, there does not exist
  $\mathcall{Z} \SUBPART \mathfrak{Y}$ such that $|\mathcall{Z}| <
  |\mathfrak{Y}|$ and \ELEMENTS{\mathcall{Z}} is correlated with
  respect to $R$.  Consider any $\tilde{\mathcall{Z}} \SUBPART
  \mathfrak{Y}$ such that \ELEMENTS{\tilde{\mathcall{Z}}} is
  correlated with respect to $R$.  As $|\tilde{\mathcall{Z}}| =
  |\mathfrak{Y}|$, every element of $\mathfrak{Y}$ is a superset of
  precisely one element of $\tilde{\mathcall{Z}}$.

  First assume all elements of $\tilde{\mathcall{Z}}$ are singletons.
  In this case no proper subset of \ELEMENTS{\tilde{\mathcall{Z}}} is
  correlated with respect to $R$.  Suppose the contrary, namely that
  some $\mathcall{W} \subset \ELEMENTS{\tilde{\mathcall{Z}}}$ is
  correlated with respect to $R$ and deduce there is some
  $\mathcall{Z}'' \SUBPART \mathfrak{Y}$ such that $\mathcall{W} =
  \ELEMENTS{\mathcall{Z}''}$, thus $|\mathcall{Z}''| < |\mathfrak{Y}|$,
  such that \ELEMENTS{\mathcall{Z}''} is correlated with respect to
  $R$.  Since no proper subset of \ELEMENTS{\tilde{\mathcall{Z}}} is
  correlated with respect to $R$, \ELEMENTS{\tilde{\mathcall{Z}}} is a
  mincor with respect to $R$ and we are done with the proof.

  Now assume not all elements of $\tilde{\mathcall{Z}}$ are singletons.
  It trivially follows there exists a minimal set
  $\widehat{\mathcall{Z}} \SUBPART \tilde{\mathcall{Z}}$ such that
  $|\widehat{\mathcall{Z}}| = |\tilde{\mathcall{Z}}|$ (thus
  $|\widehat{\mathcall{Z}}| = |\mathfrak{Y}|$) such that
  \ELEMENTS{\tilde{\mathcall{Z}}} is correlated with respect to $R$.
  \BOX

%% file: independence.tex
\section{Results on Independent Partitions}
\label{sec:independence}

This section provides important auxiliary results concerning
independent partitions.  In subsection~\ref{subsec:corrind} we
investigate the connection between independence and self-correlation.
In subsection~\ref{subsec:meetind} we prove the meet of independent
partitions is an independent partition.

\subsection{Independence and the mincor family}
\label{subsec:corrind}

The following lemma establishes that partition independence is
preserved under removal of attributes. 
\begin{lemma}
  \label{lemma3}  
  $\forall \mathfrak{Y} \in \INDPAR{\mathcall{S}} \; 
    \forall \mathfrak{X} \SUBPART \mathfrak{Y}:
    \mathfrak{X} \in \INDPART{\ELEMENTS{\mathfrak{X}}}{\PROJ{R}{\ELEMENTS{\mathfrak{X}}}}$.
\end{lemma}

\noindent
\emph{Proof:} \, 
Let $Q = \PROJ{R}{\ELEMENTS{\mathfrak{X}}}$.  We prove that $Q =
\underset{{Z} \in \mathfrak{X}}{\JOIN} \, ( \PROJ{Q}{Z} )$.  In one
direction, $Q \subseteq \underset{{Z} \in \mathfrak{X}}{\JOIN} \, (
\PROJ{Q}{Z})$ follows immediately from the definitions of relation
join and projection.  In the other direction, consider any tuple $t$
in $\underset{{Z} \in \mathfrak{X}}{\JOIN} \, (\PROJ{Q}{{Z}})$.  Let
$v$ be any tuple in $\underset{{Z} \in \mathfrak{Y}}{\JOIN} \,
(\PROJ{R}{{Z}})$ such that $t = \restr{v}{\ELEMENTS{\mathfrak{X}}}$.
But $v \in R$ because $\mathfrak{Y}$ is independent and thus $R =
\underset{{Z} \in \mathfrak{Y}}{\JOIN} \, (\PROJ{R}{{Z}})$.  As $v \in
R$, it follows that $\restr{v}{\ELEMENTS{\mathfrak{X}}} \in Q$.  But
$\restr{v}{\ELEMENTS{\mathfrak{X}}}$ is $t$, therefore $t \in Q$, and
so $\underset{{Z} \in \mathfrak{X}}{\JOIN} \, (\PROJ{Q}{{Z}})
\subseteq Q$.  \BOX

\medskip
\noindent
The next lemma is pivotal.  It shows that the mincors respect
independent partitions, in the sense that no mincor can intersect 
more than one element of an independent partition.
\begin{lemma}
  \label{lemma2}
  $\forall \mathfrak{Y} \in \INDPAR{\mathcall{S}} \; \; \forall \mathcall{W} \in \MC{R} \; \;
   \exists \mathcall{Y} \in
  \mathfrak{Y} : \mathcall{W} \subseteq \mathcall{Y}$.
\end{lemma}

\noindent
\emph{Proof:} \,
  Assume the contrary.  Then there is a mincor $\mathcall{W}$ that has
  nonempty intersection with more than one set from $\mathfrak{Y}$.
  Suppose $\mathcall{W}$ has nonempty intersection with precisely $t$
  sets from $\mathfrak{Y}$ for some $t$ such that $2 \leq t \leq q$.
  Let $\mathcall{Y}_1$, $\mathcall{Y}_2$, \ldots, $\mathcall{Y}_t$ be
  precisely those sets from $\mathfrak{Y}$ that have nonempty
  intersection with $\mathcall{W}$.  Let $\mathcall{W}_i = \mathcall{W}
  \cap \mathcall{Y}_i$, for $1 \leq i \leq t$.  Clearly,
  $\bigcup_{i=1}^t \mathcall{W}_i = \mathcall{W}$.  By
  Lemma~\ref{lemma3}:
  \[
  \PROJ{R}{\mathcall{W}} = \underset{1 \leq i \leq t}{\JOIN}
  \PROJ{R}{\mathcall{W}_i}
  \]
  Every $\mathcall{W}_i$ is a proper subset of $\mathcall{W}$.  But
  $\mathcall{W}$ is a minimal correlated set.  That implies $\neg
  \newCOR{\mathcall{W}_i}$, for $1 \leq i \leq t$.  Apply
  Lemma~\ref{lemma1} to conclude that $\displaystyle
  \PROJ{R}{\mathcall{W}_i} = \, \underset{x \in \mathcall{W}_i}{\JOIN}
  \, \PROJ{R}{\{x\}}$.  Then,
  \[
  \PROJ{R}{\mathcall{W}} = \underset{1 \leq i \leq t}{\JOIN} \; \;
  \underset{x \in \mathcall{W}_i}{\JOIN} \, \PROJ{R}{\{x\}}
  \]
  Obviously, $\displaystyle \underset{1 \leq i \leq t}{\JOIN} \; \;
  \underset{x \in \mathcall{W}_i}{\JOIN} \PROJ{R}{\{x\}} = \underset{x
    \in \mathcall{W}}{\JOIN} \PROJ{R}{\{x\}} $.  Then,
  $\PROJ{R}{\mathcall{W}} = \underset{x \in \mathcall{W}}{\JOIN} \,
  \PROJ{R}{\{x\}}$.  By Lemma~\ref{lemma1} that implies $\neg
  \newCOR{\mathcall{W}}$.  \BOX

  Furthermore, merging mincors also yields sets that respect
  independent partitions.
\begin{corollary}
  \label{cor1}
 $\forall \mathfrak{Y} \in \INDPAR{\mathcall{S}}:
  \CCMF{R} \REFINES \mathfrak{Y}$.
\end{corollary}

\noindent
\emph{Proof:} \,
  Assume the contrary.  Then for some $R$ on $\mathcall{S}$ and
  $\mathfrak{Y} \in \INDPAR{S}$:
  \[
  \exists \mathcall{X} \in \CCMF{R} \;\; \forall \mathcall{Y} \in
  \mathfrak{Y} \;\; \exists A \in \mathcall{X} : A \not\in \mathcall{Y}
  \]
  First note that $\mathcall{X}$ is not a singleton, otherwise
  $\mathcall{X}$ would be contained in some set from $\mathfrak{Y}$.
  So, $|\mathcall{X}| \geq 2$ and according to Definition~\ref{def4},
  $\mathcall{X}$ is the union of one or more mincors, each of size
  $\geq 2$, and $\mathcall{X}$ is connected.  But by assumption
  $\mathcall{X}$ is not a subset of any set from $\mathfrak{Y}$ and so
  there has to be some mincor $\mathcall{W} \in \mathcall{X}$ that has
  nonempty intersection with at least two sets from $\mathfrak{Y}$ .
  However, that contradicts Lemma~\ref{lemma2}.  \BOX

\medskip
\noindent
Note that \label{example1} $\CCMF{R}$ is not necessarily an
independent partition.  For example, consider $R'$ defined in
(\ref{eq3}) \vpageref{eq3}.
%
As explained \vpageref{example2}, $\MF{R'} = \{ \{A, B\}, \{C, D\} \}$
and thus $\CCMF{R'} = \{ \{A, B\}, \{C, D\} \}$, too.  But $\{ \{A,
B\}, \{C, D\} \}$ is not an independent partition with respect to
$R'$.  In fact, there is no independent partition of $\mathcall{S}$
except for the trivial partition as $|R'|$ is a prime number.

Now consider another relation $R''$ on the same scheme:
\begin{align*}
R'' = \{ & \LA a_1 b_1  c_1 d_1 \RA, 
             \LA a_1 b_1  c_1 d_2 \RA,
             \LA a_1 b_1 c_2 d_2 \RA, 
             \LA a_1 b_2 c_1 d_1 \RA,
             \LA a_1 b_2 c_1 d_2 \RA, 
             \LA a_1 b_2 c_2 d_2 \RA, \\
&             \LA a_2 b_2 c_1 d_1 \RA,
             \LA a_2 b_2 c_1 d_2\RA,
             \LA a_2 b_2 c_2 d_2 \RA 
              \}
\end{align*}
But $\MF{R''} = \{ \{A, B\}, \{C, D\} \} = \CCMF{R''}$ just as in the
case of $R'$.  Now $\{ \{A, B\}, \{C, D\} \}$ is an independent
partition with respect to $R''$ because $R'' = \PROJ{R''}{\{A,B\}}
\JOIN \PROJ{R''}{\{C,D\}}$.

So, in the case of $R''$, the connected components of the mincor
family constitute an independent partition, while that is not true for
$R'$, although the mincor families of both relations are the same.  We
conclude that computing the mincor family does not suffice to obtain
an independent partition.  Therefore, we use a more involved approach
in which the computation of the mincor family is but the first step
towards the computation of the maximum independent partition.

\subsection{The meet of independent partitions}
\label{subsec:meetind}

The following lemma allows us to define the maximum
independent partition as the meet of all independent partitions.
\begin{lemma}
\label{lemma:meetindependent}
$\forall \mathfrak{X}, \mathfrak{Y} \in \INDPAR{\mathcall{S}}: \mathfrak{X} \GLB \mathfrak{Y} \in \INDPAR{\mathcall{S}}$.
\end{lemma}

\noindent
\emph{Proof: (sketch)} \,
  Let $\mathfrak{X}, \mathfrak{Y} \in \INDPAR{\mathcall{S}}$.
  We assume $\mathfrak{X} \LUB \mathfrak{Y}$ is connected.  There is
  no true loss of generality in that because the proof below can be
  done componentwise if $\mathfrak{X} \LUB \mathfrak{Y}$ is not
  connected.  Relative to an arbitrary element of $\mathfrak{X}$, 
  say~$X_1$, we define the family $\mathfrak{Z} = \{Z_0, Z_1 \LDOTS Z_k\}$
  over~$\mathcall{S}$ as follows.  $\mathfrak{Z}$~is a partition 
  of~$\mathcall{S}$ and its elements are constructed in an ascending order
  of the index according to the following rule:
  \[
    Z_i =
    \begin{cases}
      X_1, & \text{ if } i = 0 \\
      \bigcup \{A \setminus Z_{i-1} \, | \,  A \in \mathfrak{Y} \wedge A \cap  Z_{i-1} \not= \emptyset \}, & \text{ if } i \text{ is odd} \\
      \bigcup \{A \setminus Z_{i-1} \, | \,  A \in \mathfrak{X} \wedge A \cap  Z_{i-1} \not= \emptyset \}, & \text{ if } i \text{ is even and } i > 0
    \end{cases}
  \]
  Let us define $B_i = \left\{ \bigcup_{j=0}^i Z_j \right\} \GLB
  \mathfrak{X} \GLB \mathfrak{Y}$ for $0 \leq i \leq k$.  Clearly,
  $B_0 = \{X_1\} \GLB \mathfrak{Y}$, $B_i = B_{i-1} \origcup (\{ Z_i
  \} \GLB \mathfrak{X} \GLB \mathfrak{Y})$ for $1 \leq i \leq k$ and
  $B_k = \mathfrak{X} \GLB \mathfrak{Y}$.  Furthermore, $\ELEMENTS{B_k} =
  \mathcall{S}$ and thus $\textstyle \PROJ{R}{\ELEMENTS{B_k}} = R$.
We prove by induction on~$i$ that for all~$i$ such that $0 \leq i \leq k$:
  \begin{equation}
    \label{eq8}
    \textstyle \PROJ{R}{\ELEMENTS{B_i}} = \underset{C \in B_i}{\JOIN}\PROJ{R}{C}
  \end{equation}
and hence the result follows.
  
\medskip
\noindent
\emph{Basis.} Let $i=0$.  Let the elements of $\mathfrak{Y}$ that have
nonempty intersection with $X_1$ be called $Y_1$, \ldots, $Y_j$.
Obviously, there is at least one of them.  The claim is that
$\PROJ{R}{X_1} = \JOIN_{i=1}^j \PROJ{R}{(X_1 \origcap Y_i)}$.
That follows immediately from Lemma~\ref{lemma3}.

\medskip
\noindent
\emph{Inductive Step.}  Assume the claim holds for some $B_{i-1}$ such
that $0 \leq i-1 < k$ and consider $B_i$.  As already mentioned, $B_i
= B_{i-1} \cup (\{ Z_i \} \GLB \mathfrak{X} \GLB \mathfrak{Y})$.

Without loss of generality, assume $i$ is odd.  Very informally
speaking, $Z_i$ is the union of some elements of $\mathfrak{Y}$ that
overlap with some elements (from $\mathfrak{X}$) in $B_{i-1}$, minus
the overlap.  Therefore, we can write $B_i = B_{i-1} \cup (\{ Z_i \}
\GLB \mathfrak{X})$ because under the current assumption, it is
$\mathfrak{X}$ rather than $\mathfrak{Y}$ that dictates the grouping
together of the elements of $Z_i$ in $B_i$.  More specifically, since
$i \not= k$, there are elements from $\mathfrak{X}$ whose elements do
not appear in the current $B_i$; those elements of $\mathfrak{X}$
dictate the aforementioned grouping.

So, $B_i$ is the union of two disjoint sets whose elements are from
$\mathfrak{X} \GLB \mathfrak{Y}$, namely $B_{i-1}$ and $\{ Z_i \} \GLB
\mathfrak{X}$.  By the inductive hypothesis,
$\PROJ{R}{\ELEMENTS{B_{i-1}}} = \underset{C \in
  B_{i-1}}{\JOIN}\PROJ{R}{C}$.

Consider $\{ Z_i \} \GLB \mathfrak{X}$ and call its elements, $T_1$,
\ldots, $T_m$.  Without loss of generality, consider $T_1$.  Our
immediate goal is to prove that $\PROJ{R}{( \, (\ELEMENTS{B_{i-1}} )
  \, \cup T_1)} = \underset{C \in B_{i-1} \cup
  \{T_1\}}{\JOIN}\PROJ{R}{C}$.  Note that $T_1$ is a subset of some
$Y' \in \mathfrak{Y}$ such that $Y'$ has nonempty intersection with
\ELEMENTS{B_{i-1}}, $T_1$ itself being disjoint with $B_{i-1}$.
Furthermore, $T_1$ is the intersection of $Y'$ with some $X' \in
\mathfrak{X}$.  $X'$~is disjoint with
\ELEMENTS{B_{i-1}}, otherwise the elements of~$T_1$ would be part of
\ELEMENTS{B_{i-1}}.  Furthermore, every element of $B_{i-1}$ is a
subset of some element of $\mathfrak{X}$ that is not $X'$.  Let the
elements of $\mathfrak{X}$ that have subsets-elements of $B_{i-1}$ be
$X_1$, \ldots, $X_p$.  Note that $X_1 \cup \cdots \cup X_p =
\ELEMENTS{B_{i-1}}$.  By Lemma~\ref{lemma3}, it is the case that
\begin{equation}
\label{eq9}
\PROJ{R}{(X_1 \cup \cdots \cup X_p \cup T_1)} = 
\PROJ{R}{X_1} \JOIN \cdots \JOIN \PROJ{R}{X_p} \JOIN \PROJ{R}{T_1} 
\end{equation}
since $T_1$ is a subset of $X'$ and $X'$ is none of $X_1$, \ldots,
$X_p$.  However, $X_1 \cup \cdots \cup X_p \cup T_1 =
(\ELEMENTS{B_{i-1}}) \, \cup T_1$ by an earlier observation and
$\PROJ{R}{X_1} \JOIN \cdots \JOIN \PROJ{R}{X_p} = \underset{C \in
  B_{i-1}}{\JOIN}\PROJ{R}{C}$.  Substitute that in equation~\ref{eq9}
to obtain
\begin{equation}
\label{eq10}
\PROJ{R}{(\ELEMENTS{B_{i-1}} \cup T_1)} = \left( 
  \underset{C
    \in B_{i-1}}{\JOIN}\PROJ{R}{C} \right)
\JOIN \PROJ{R}{T_1} =
\underset{C \in B_{i-1} \cup
  \{T_1\}}{\JOIN}\PROJ{R}{C}
\end{equation}
which is what we wanted to prove with respect to $T_1$.

\medskip
\noindent
We can use (\ref{eq10}) as the basis of a nested induction.  More
specifically, we prove that
\[
\PROJ{R}{\, ((\ELEMENTS{B_{i-1}} ) \, \cup T_1 \cup \cdots \cup T_k)} = 
\left(\underset{C
  \in B_{i-1}}{\JOIN}\PROJ{R}{C} \right) \JOIN \PROJ{R}{T_1} \JOIN \cdots \JOIN \PROJ{R}{T_k}
\]
implies
\[
\PROJ{R}{\, ((\ELEMENTS{B_{i-1}} ) \, \cup T_1 \cup \cdots \cup T_{k+1})} = 
\left(\underset{C
  \in B_{i-1}}{\JOIN}\PROJ{R}{C} \right) \JOIN \PROJ{R}{T_1} \JOIN \cdots \JOIN \PROJ{R}{T_{k+1}}
\]
for any $k \in \{1, 2, \ldots, m-1\}$.  The nested induction can be
proved in a straightforward manner, having in mind the proof of
(\ref{eq10}).  That implies the desired:
\[
\PROJ{R}{\, ((\ELEMENTS{B_{i-1}} ) \, \cup T_1 \cup \cdots \cup T_m)} = 
\left(\underset{C
  \in B_{i-1}}{\JOIN}\PROJ{R}{C} \right) \JOIN \PROJ{R}{T_1} \JOIN \cdots \JOIN \PROJ{R}{T_m}
\]
And that concludes the proof because $\ELEMENTS{B_i} =
\ELEMENTS{B_{i-1}} \cup T_1 \cup \cdots \cup T_m$.
\BOX \\
The proof of Lemma~\ref{lemma:meetindependent} relies on the fact that
all sets we consider are finite.

\medskip
\noindent
As a corollary of Lemma~\ref{lemma:meetindependent}, the maximum
independent partition, which is the object of our study, is
well-defined: $\GLB\, \INDPAR{\mathcall{S}}$ exists, it is unique, and
is an element of $\INDPAR{\mathcall{S}}$.  For notational convenience 
we introduce another term for that object.  We say that $\GLB\,
\INDPART{\mathcall{S}}{R}$ is the \emph{focus} of~$R$ and denote
it by \FOC{R}.  A trivial observation is that \label{prop4}
\INDPART{\mathcall{S}}{R} coincides with \FILTER{\FOC{R}}.

%% file: main.tex
\section{A Fixed Point Characterisation of the Maximum Independent
  Partition}
\label{sec:main}

In this section we identify the object of our study as the least fixed
point of~$\alpha$, where $\alpha$ is a transformer on the lattice of
all partitions of $S$.  Furthermore, we present an
iterative fixed point approximation procedure for computing the
maximum independent partition.

\subsection{Function $\alpha$}

First we introduce a helper function.  Let $A$ be a ground set.  The
function $\xi$ maps superfamilies over $A$ to families over $A$ as
follows.  For any superfamily $\mathfrak{F}$:
  \[
  \xi(\mathfrak{F}) \stackrel{\text{\tiny def}}{=} \bigl\{ \origcup Z
  | \, Z \in \mathfrak{F} \bigr\}
  \]
Syntactically speaking, $\xi$ removes the innermost pairs of
parentheses.  For instance, suppose $A  = \{a,b,c,d\}$ and
$\mathfrak{F} = \{ \{\{a\}, \{b,c\} \}, \{\{d\} \} \}$.
Then $\xi(\mathfrak{F}) = \{ \{a, b ,c \}, \{d\} \}$.

We now define the central function of the present study. It takes a
partition of $S$, identifies the mincors of the corresponding quotient
relation, merges the overlapping mincors, and uses~$\xi$ to map the
result back to a partition of $S$.
\begin{definition}[function $\alpha$]
  \label{def:alpha}
  $\alpha_R: \PART{\mathcall{S}} \rightarrow \PART{\mathcall{S}}$,
  shortly $\alpha$ when $R$ is understood, is defined as follows for
  any $\mathfrak{X} \in \PART{\mathcall{S}}$:
  \[
 \alpha_R(\mathfrak{X}) 
    \stackrel{\text{def}}{=} \xi(\CCMF{\QUOT{R}{\mathfrak{X}}})  \]
\end{definition}

Notably, $\alpha$ is \emph{not monotone} 
\label{alpha-not-monotone} in general as demonstrated by
the following example.  Let $\widetilde{\mathcall{S}} = \{A, \AB B, \AB
C, \AB D, \AB E\}$ and let each attribute have precisely two values,
say $A = \{a_1, a_2\}$ and so on.  Let $Q$ be the relation
obtained from the complete relation over $\widetilde{\mathcall{S}}$
after deleting all tuples containing $a_1 b_1 c_1$, all tuples
containing $d_2 e_2$, and the tuples $\LA a_2 b_1 c_1 d_2 e_1\RA, \LA
a_2 b_2 c_1 d_2 e_1 \RA$.  In other words,
\begin{xxalignat}{2}
  &Q = 
  && \hspace{-3mm} \{
 \LA  a_1 b_1 c_2 d_1 e_1 \RA,
  \LA  a_1 b_1 c_2 d_1 e_2 \RA,
  \LA  a_1 b_1 c_2 d_2 e_1 \RA,
  \LA  a_1 b_2 c_1 d_1 e_1 \RA,
  \LA  a_1 b_2 c_1 d_1 e_2 \RA,
  \LA  a_1 b_2 c_1 d_2 e_1 \RA \\
&&& \hspace{-1mm} \LA  a_1 b_2 c_2 d_1 e_1 \RA,
  \LA  a_1 b_2 c_2 d_1 e_2 \RA,
  \LA  a_1 b_2 c_2 d_2 e_1 \RA,
  \LA  a_2 b_1 c_1 d_1 e_1 \RA,
  \LA  a_2 b_1 c_1 d_1 e_2 \RA,
  \LA  a_2 b_1 c_2 d_1 e_1 \RA \\
&&& \hspace{-1mm} \LA  a_2 b_1 c_2 d_1 e_2 \RA,
  \LA  a_2 b_1 c_2 d_2 e_1 \RA,
  \LA  a_2 b_2 c_1 d_1 e_1 \RA,
  \LA  a_2 b_2 c_1 d_1 e_2 \RA,
  \LA  a_2 b_2 c_2 d_1 e_1 \RA,
  \LA  a_2 b_2 c_2 d_1 e_2 \RA,
  \LA  a_2 b_2 c_2 d_2 e_1 \RA 
  \} 
\end{xxalignat}

Let us see which sets of attributes are self-correlated with respect
to $Q$.  The only two-element subset of $\widetilde{\mathcall{S}}$ that
is self-correlated is $\{D, E\}$.  Further, $\{A, B, C\}$ is
self-correlated.  It follows $\MF{Q} = \{ \{A, B, C\}, \AB \{D, E\} \}$.
Consider the following two partitions of $\widetilde{\mathcall{S}}$:
$\mathfrak{X}_1 = \{ \{A\}, \{B\}, \{C\}, \{D\}, \{E\} \}$ and
$\mathfrak{X}_2 = \{ \{A\}, \{B, D\}, \{C, E\} \}$.  Obviously,
$\mathfrak{X}_1 \REFINES \mathfrak{X}_2$.  It is clear that
$\alpha(\mathfrak{X}_1) = \{ \{A, B, C\}, \{D, E\} \}$.  Consider
$\alpha(\mathfrak{X}_2)$.  The set $\{ \{B, D\}, \{C, E\} \}$ is
self-correlated because of the lack of $\LA b_1, d_2\RA$ and $\LA c_1,
e_2 \RA$ in any tuple, which in its turn is due to the fact that
$d_2$ and $e_2$ do not occur in any tuple of $R$.  The sets $\{ \{A\},
\{B, D\} \}$ and $\{ \{A\}, \{C, E\} \}$ are uncorrelated.  It follows
that $\alpha(\mathfrak{X}_2) = \{ \{A\}, \{B, C, D, E\} \}$, and thus
$\alpha(\mathfrak{X}_1) \not\REFINES \alpha(\mathfrak{X}_2)$.

\medskip
\noindent
However, we have the following property of~$\alpha$ that shall later
be exploited.
\begin{proposition}
  \label{prop3} 
  $\alpha$ is an inflationary function on $(\PART{\mathcall{S}},
  \REFINES)$.
\end{proposition}

\subsection{Independence and function $\alpha$}

The following central result establishes that the independent
partitions are precisely the fixed points of~$\alpha$.
\begin{theorem}
  \label{theorem:alphaindependent}
  $\forall \mathfrak{X} \in \PART{\mathcall{S}}: 
  \mathfrak{X} \in \INDPAR{\mathcall{S}} \leftrightarrow \alpha(\mathfrak{X}) =
  \mathfrak{X}$.
\end{theorem}

\noindent
\emph{Proof:} \,
  In one direction, assume $\mathfrak{X} \in \INDPAR{\mathcall{S}}$.
  \QUOT{R}{\mathfrak{X}} is complete by Proposition~\ref{prop2}.  By
  definition, that is $\QUOT{R}{\mathfrak{X}} = \vartimes_{Y \in
    \mathfrak{X}} Y$.  By the definition of $\bowtie$,
  $\PROJ{(\QUOT{R}{\mathfrak{X}})}{\mathfrak{X}} = {\JOIN}_{{Y \in
      \mathfrak{X}}} \PROJ{(\QUOT{R}{\mathfrak{X}})}{\{ Y \}}$.  It
  follows that $\neg \newCOR{\mathfrak{X}}$ by Lemma~\ref{lemma1}.
  So, $\MC{\QUOT{R}{\mathfrak{X}}} = \emptyset$ and
  $\MF{\QUOT{R}{\mathfrak{X}}} = \SNGL{\QUOT{R}{\mathfrak{X}}}$ by
  Definition~\ref{def4}.  Then $\CCMF{\QUOT{R}{\mathfrak{X}}} = \{ \{A
  \} \, | \, A \in \mathfrak{X} \}$.  Therefore,
  $\xi(\CCMF{\QUOT{R}{\mathfrak{X}}}) = \{ A \, | \, A \in
  \mathfrak{X} \} = \mathfrak{X}$.  But
  $\xi(\CCMF{\QUOT{R}{\mathfrak{X}}})$ is $\alpha(\mathfrak{X})$ by
  definition.  Therefore, $\alpha(\mathfrak{X}) = \mathfrak{X}$.
  
In the other direction, assume $\alpha(\mathfrak{X}) = \mathfrak{X}$.
That is, $\xi(\CCMF{\QUOT{R}{\mathfrak{X}}}) = \mathfrak{X}$, which in
its turn implies $\CCMF{\QUOT{R}{\mathfrak{X}}} = \{ \{A \} \, | \, A
\in \mathfrak{X} \}$ because $\CCMF{\QUOT{R}{\mathfrak{X}}}$ is a
superfamily such that every element from $\mathcall{S}$ is in precisely
one element of precisely one element of it.  The remainder of the
proof mirrors the above one. 
\BOX

Having in mind the observation \vpageref{prop4} that
\INDPART{\mathcall{S}}{R} coincides with \FILTER{\FOC{R}}, we derive
the following corollary of Theorem~\ref{theorem:alphaindependent}.
\begin{corollary}
  \label{corollary:alphaindependent}
  \FILTER{\FOC{R}} is closed with respect to $\alpha$.
\end{corollary}

\medskip
\noindent
The following lemma says that the mincors of a quotient relation
respect the focus of the relation in the sense that for every
mincor of~\QUOT{R}{\mathfrak{X}}, the union of its elements is a
subset of some element of the focus.
\begin{lemma}
  \label{lemma4}
 $\forall \mathfrak{X} \in  \IDEAL{\FOC{R}} \;   
    \forall {T} \in \MC{\QUOT{R}{\mathfrak{X}}} \;
    \exists \mathcall{Y} \in \FOC{R} : \ELEMENTS{{T}} \subseteq \mathcall{Y}$.
\end{lemma}

\noindent
\emph{Proof:} \,
  Assume the contrary.  That is, for some partition $\mathfrak{X}$
  that refines the focus there is a mincor ${T}$ of
  \QUOT{R}{\mathfrak{X}} such that $\ELEMENTS{{T}}$ has
  nonempty intersection with at least two subsets, call them
  $\mathcall{Y}_1$ and $\mathcall{Y}_2$, of the focus.  Use
  Lemma~\ref{lemma:mincors} to conclude there is some $\mathcall{Z}
  \SUBPART {T}$ such that $|\mathcall{Z}| = |{T}|$
  and $\ELEMENTS{\mathcall{Z}} \in \MC{R}$.  Since $|\mathcall{Z}| =
  |{T}|$, it must be the case that \ELEMENTS{\mathcall{Z}}
  has nonempty intersection with both $\mathcall{Y}_1$ and
  $\mathcall{Y}_2$.  But the focus is an independent partition.  We
  derived that a mincor of $R$, namely \ELEMENTS{\mathcall{Z}},
  intersects two distinct elements of an independent partition.  That
  contradicts Lemma~\ref{lemma2} directly.
\BOX

\medskip
\noindent
We already established (see Proposition~\ref{prop3}) that~$\alpha$ is
an inflationary function.  The next lemma, however,
establishes a certain restriction: the application of~$\alpha$ on a
dependent partition can yield another dependent partition or at most
the focus, and never an independent partition ``above'' the focus.
\begin{lemma}
  \label{lemma:idealclosed}
  \IDEAL{\FOC{R}} is closed with respect to~$\alpha$.
\end{lemma}

\noindent
\emph{Proof:} \,
  We prove that $\forall \mathfrak{X} \in \IDEAL{\FOC{R}}:
  \alpha(\mathfrak{X}) \REFINES \FOC{R}$.  Recall that
  $\alpha(\mathfrak{X})$ is a partition of $\mathcall{S}$ and it
  abstracts $\mathfrak{X}$.  Assume the claim is false.  Then there is
  a partition $\mathfrak{X}$ such that $\mathfrak{X} \REFINES
  \IDEAL{\FOC{R}}$ but $\alpha(\mathfrak{X}) \not\REFINES
  \IDEAL{\FOC{R}}$.  Then there is
  some $\mathcall{P} \in \alpha(\mathfrak{X})$ such that
  $\mathcall{P}$ has nonempty intersection with at least two elements,
  call them $\mathcall{Y}_1$ and $\mathcall{Y}_2$, of \FOC{R}.  However,
  $\mathcall{P}$ is $\xi({C})$ for some ${C}$ that is
  a connected component---relative to the ground set
  $\mathfrak{X}$---of the mincor family of \QUOT{R}{\mathfrak{X}}.
  Consider ${C}$.  It is the union of one or more mincors of
  \QUOT{R}{\mathfrak{X}}, those mincors being subsets of
  $\mathfrak{X}$.

  Since $\mathfrak{X} \REFINES \FOC{R}$, no element of $\mathfrak{X}$
  can intersect both $\mathcall{Y}_1$ and $\mathcall{Y}_2$.  It follows
  that at least one mincor $\mathcall{M} \in {C}$ is such that
  \ELEMENTS{\mathcall{M}} intersects both $\mathcall{Y}_1$ and
  $\mathcall{Y}_2$.  But that contradicts Lemma~\ref{lemma4}.
  \BOX

\medskip
\noindent
The next and final central result allows us to compute the focus of~$R$ 
by an iterative application of~$\alpha$, starting with the partition into
singletons.
\begin{theorem}
  \label{theorem:main}
  For some $m$ such that $1 \leq m \leq |\mathcall{S}|$, $\alpha^m(\bot) = \FOC{R}$.
\end{theorem}

\noindent
\emph{Proof:} \, Consider the sequence:
\[
{C} = \bot,\ \alpha(\bot),\ \alpha^2(\bot),\ \ldots
\]
It is a chain in the lattice $(\PART{\mathcall{S}}, \REFINES)$,
as $\alpha(\mathfrak{X})$ abstracts $\mathfrak{X}$ for all
$\mathfrak{X}$ (see Proposition~\ref{prop3}), therefore all those
elements are comparable with respect to $\REFINES$.  ${C}$
has only a finite number of distinct elements as the said lattice is
finite.

First note that every element of ${C}$ is in
$\IDEAL{\FOC{R}}$.  Indeed, assuming the opposite immediately
contradicts Lemma~\ref{lemma:idealclosed}.

Then note that for every $\mathfrak{X} \in \IDEAL{\FOC{R}} \setminus
\{\FOC{R} \}$, it is the case that $\alpha(\mathfrak{X}) \not=
\mathfrak{X}$.  Assuming the opposite implies $\mathfrak{X}$ is a
fixed point of $\alpha$, contradicting
Corollary~\ref{corollary:alphaindependent}.  Proposition~\ref{prop3}
implies a stronger fact: for every $\mathfrak{X} \in \IDEAL{\FOC{R}}
\setminus \{\FOC{R} \}$, it is the case that $\mathfrak{X}
\REFINESproperly \alpha(\mathfrak{X})$.  But $\IDEAL{\FOC{R}}$ is a
finite lattice.  It follows immediately that for some value $m$ not
greater than $|\mathcall{S}|$, $\alpha^m(\bot)$ equals the top
of $\IDEAL{\FOC{R}}$, \emph{viz.} $\FOC{R}$.
\BOX

\medskip
\noindent
We thus obtain Kleene's iterative least fixed point approximation
procedure~\cite{LNS}, however for inflationary functions instead of
monotone ones.
\begin{corollary}
  \label{cor:fixed-point-comp}
  The following algorithm:
\begin{align*}
  & \mathfrak{X} \leftarrow \bot \\
  & \textbf{while} \;  \mathfrak{X}  \not= \alpha(\mathfrak{X}) \\
  & \makebox[.93cm]{ \ } \mathfrak{X} \leftarrow \alpha(\mathfrak{X}) \\
  & \textbf{return} \; \mathfrak{X}
\end{align*}
computes the least fixed point of $\alpha$, \emph{i.e.}, the maximum
independent partition of $\mathcall{S}$ with respect to $R$. \BOX
\end{corollary}

Here is a small example illustrating the work of that algorithm.
Consider $\mathcall{S}$ and $R'$ defined in (\ref{eq3}) \vpageref{eq3}.
$\bot$ is $\{ \{A\}, \{B\}, \{C\}, \{D\} \}$.  Let us compute $\alpha(\bot)$,
that is, $\xi(\CCMF{\QUOT{R'}{ \bot }})$.  \QUOT{R'}{\bot} is
the same as \QUOT{R'}{\mathfrak{X}_2} \vpageref{eq7}, namely:
\begin{alignat*}{2}
 \QUOT{R'}{\bot} & = \{ && 
                                                                   \LA 
                                                                       \LAO a_1 \RAO  
                                                                       \LAO  b_1 \RAO 
                                                                       \LAO c_1 \RAO 
                                                                       \LAO d_1 \RAO  
                                                                   \RA, 
                                                                   \LA 
                                                                       \LAO a_1 \RAO  
                                                                       \LAO  b_1 \RAO 
                                                                       \LAO c_2 \RAO 
                                                                       \LAO d_2 \RAO  
                                                                   \RA, 
                                                                   \LA 
                                                                       \LAO a_1 \RAO  
                                                                       \LAO  b_2 \RAO 
                                                                       \LAO c_1 \RAO 
                                                                       \LAO d_2 \RAO  
                                                                   \RA, \notag \\ 
                                                      &&& 
                                                                   \LA 
                                                                       \LAO a_2 \RAO  
                                                                       \LAO  b_2 \RAO 
                                                                       \LAO c_1 \RAO 
                                                                       \LAO d_1 \RAO  
                                                                   \RA, 
                                                                   \LA 
                                                                       \LAO a_2 \RAO  
                                                                       \LAO  b_2 \RAO 
                                                                       \LAO c_2 \RAO 
                                                                       \LAO d_2 \RAO  
                                                                   \RA 
  \}
\end{alignat*}
Let us compute $\CCMF{\QUOT{R'}{ \bot }}$.  Having in mind that
$\MF{R'} = \{ \{A, B\}, \{C, D\} \}$ as explained \vpageref{example2},
conclude that $\CCMF{\QUOT{R'}{ \bot }} = \{ \{ \{A, B\}\}, \{\{C,
D\}\} \}$.  Therefore, $\xi(\CCMF{\QUOT{R'}{ \bot }}) = \{ \{ A, B\},
\{C, D\} \}$.  That differs from $\bot$ and the \textbf{while} loop
is executed again.  \QUOT{R'}{\alpha(\bot)} is
the same as \QUOT{R'}{\mathfrak{X}_1} \vpageref{eq7}, namely:
\begin{alignat*}{2}
 \QUOT{R'}{\alpha(\bot)} & = \{ & & \LA \LAO a_1 b_1 \RAO  \LAO c_1 d_1 \RAO  \RA, 
                                                      \LA \LAO a_1 b_1 \RAO  \LAO c_2 d_2 \RAO \RA, 
                                                      \LA \LAO a_1 b_2 \RAO  \LAO c_1 d_2 \RAO \RA, \notag \\ 
                                                     &&& \LA \LAO a_2 b_2 \RAO  \LAO c_1 d_1 \RAO \RA, 
                                                      \LA \LAO a_2 b_2 \RAO  \LAO c_2 d_2 \RAO \RA  \} 
\end{alignat*}
Let us compute $\CCMF{\QUOT{R'}{ \alpha(\bot) }}$.  To that end, note
that $\alpha(\bot) = \{ \{A, B\}, \{C, D\} \}$ is self-correlated with respect to
\QUOT{R'}{\{ \{A,B\}, \{C,D\} \} } because of the lack of, for
instance, both $\LA a_1, b_2 \RA$ and $\LA c_1, d_1 \RA$ in any tuple of
\QUOT{R'}{\alpha(\bot)}.  It follows that $\CCMF{\QUOT{R'}{
    \alpha(\bot) }} = \{ \{ \{A, B\}, \{C, D\}\} \}$ and, therefore,
$\alpha^2(\bot) = \xi(\CCMF{\QUOT{R'}{ \alpha(\bot) }}) = \{ \{ A, B, C, D\} \}$.  
That differs from $\alpha(\bot)$ and the \textbf{while} loop is executed once
more.  At the end of that execution, it turns out that $\alpha^3 (\bot)$
equals $\alpha^2 (\bot)$ and the algorithm terminates, returning as the result 
$\{ \{ A, B, C, D\} \}$, the trivial partition.

%% file: relatedwork.tex
\section{Related Work}
\label{sec:relatedwork}

An algorithm that factorizes a given relation into prime factors is
proposed in~\cite[algorithm \textsc{Prime Factorization}]{AKO}.  It
runs in time $O(mn\lg{n})$ where $m$ is the number of tuples and $n$
is the number of attributes.  Since $mn$ is the input size, that time
complexity is very close to the optimum.  The theoretical foundation
of \textsc{Prime Factorization} is a theorem (see~\cite[Proposition~10]{AKO}) 
that says a given relation $S$ has a
factor $F$ iff, with respect to any attribute $A$ and any value $v$ of
its domain, $F$ is a factor of both $Q$ and $R$ where $Q$ and $R$ are
relations such that $Q \cup R = S$ and $Q$ consists precisely of the
tuples in which the value of $A$ is $v$.  In other words, the approach
of~\cite{AKO} to the problem of computing the prime factors is
``horizontal splitting'' of the given relation using the selection
operation from relational algebra.  The approach of this paper to that
same problem is quite different.  We utilise ``vertical splitting'',
using the projection operation of relational algebra.  The theoretical
foundation of our approach is based on the concept of self-correlation
of a subset of the attributes; that concept has no analogue in~\cite{AKO}.

An excellent exposition of the benefits of the factorisation of
relational data is~\cite{OZ}.  The factorised representation both
saves space, where the gain can potentially be as good as exponential,
and time, speeding up the processing of information whose
un-factorised representation is too big.  \cite{AKO09}~proposes a way
of decomposing relational data that is incomplete and~\cite{Rendle}
proposes factorisation of relational data that facilitates machine
learning.

Clusterisation of multidimensional data into non-intersecting classes
called clusters is an important, hard and computationally demanding
problem.  \cite{Gunnemann}~investigates clustering in high-dimensional
data by detection of orthogonality in the latter.  \cite{Lin}~proposes
so called community discovering, which is a sort of clusterisation,
in media social networks by utilising factorisation of a relational
hypergraph.

The foundation of this paper is the work of Gurov \emph{et
  al.}~\cite{GurOstSchae11ISoLA} that investigates relational
factorisation of a restricted class of relations called there simple
families.  \cite{GurOstSchae11ISoLA} introduces the concept of
correlation between the attributes and proposes a fast and practical
algorithm that computes the optimum factorisation of a simple family
by using a subroutine for correlation.  The fundamental approach of
this paper is an extension of that, however now correlation is
considerately more involved, being not a binary relation between
attributes but a relation of arbitrary arity (this is the only place
where ``relation'' means relation in the Set Theory sense, that is, a
set of ordered tuples).

%% file: conclusions.tex
\section{Conclusion}
\label{sec:conclusion}

This paper illustrates the utility of fixed points to formally express
maximum independence in relations by means of minimum correlated sets
of attributes.  By using minimum correlated sets, we define an
inflationary transformer over a finite lattice and show the maximum
independent partition is the least fixed point of this transformer.
Then we prove the downward closure of that least fixed point is closed
under the transformer. Hence, the least fixed point can be computed by
applying the transformer iteratively from the bottom element of the
lattice until stabilization. This iterative construction is the same
as Kleene's construction, but does not rely on monotonicity of the
transformer to guarantee that it computes the least fixed point.

A topic for future work is 
to introduce a quantitative measure for the degree of independence
between sets of attributes and investigate approximate relational
factorisation.

\paragraph{Acknowledgement}
We are indebted to Zolt\'{a}n \'{E}sik for pointing out to us that the
CPO Fixpoint Theorem~III of~\cite[pp.\ 188]{DaveyPriestley} about the
existence of least fixed points of inflationary functions (called
there ``increasing functions'') in CPOs does in fact not hold, and to
Valentin Goranko for directing us to the work by Olteanu et al.
Finally, we thank the reviewers of this paper for the thorough
assessments and the valuable suggestions that allowed us
to improve the quality of the presentation.

%% file: thebibliography.tex
\addcontentsline{toc}{chapter}{References}
\bibliographystyle{eptcs}
\bibliography{cartfact}